\documentclass[conference]{IEEEtran}
\IEEEoverridecommandlockouts
\usepackage{cite}
\usepackage{amsmath,amssymb,amsfonts}
\usepackage{algorithm}
\usepackage{algpseudocode}
\usepackage{graphicx}
\usepackage{textcomp}
\usepackage{xcolor}
\usepackage{fancyhdr}
\usepackage{caption,subcaption}
\usepackage[bookmarks=true,breaklinks=true,letterpaper=true,colorlinks,linkcolor=black,citecolor=blue,urlcolor=black]{hyperref}
\usepackage{listings}
\usepackage[export]{adjustbox}
\usepackage{multirow}
\usepackage{booktabs}
\usepackage{stfloats}

\def\BibTeX{{\rm B\kern-.05em{\sc i\kern-.025em b}\kern-.08em
    T\kern-.1667em\lower.7ex\hbox{E}\kern-.125emX}}

\fancypagestyle{firstpage}{
  \fancyhf{}

  \fancyhead[C]{}
  \fancyfoot[C]{\thepage}
}

\title{ Per-Bank Memory Bandwidth Regulation for Predictable and Performant Real-Time Systems
}

\author{\IEEEauthorblockN{Connor Rudy Sullivan, Amin Mamandipoor, Cole Ridge Strickler, Heechul Yun}
\IEEEauthorblockA{\textit{Electrical Engineering and Computer Science} \\
\textit{The University of Kansas}\\
Lawrence, KS, USA \\
\{connor.sullivan13, aminm, colestrickler, heechul.yun\}@ku.edu}
}

\begin{document}
\maketitle
\thispagestyle{firstpage}
\begin{abstract}
Modern multicore system-on-chips (SoCs) share off-chip DRAM across cores, where bank-level interference can significantly degrade performance and threaten real-time guarantees. While prior work has focused on per-core bandwidth regulation, these approaches treat main memory as a monolithic resource and overlook DRAM's inherent bank-level parallelism. 

We show that DRAM interference is fundamentally a bank-level phenomenon. We characterize the guaranteed bandwidth of modern DRAM, demonstrate that it remains effectively constant across generations, and show how this limitation can be exploited by single-bank attacks. These results highlight the need for bank-aware memory management for predictable and efficient real-time systems.

We design and implement a novel per-bank memory bandwidth regulator in an open-source RISC-V SoC and evaluate it using FireSim with both synthetic and real-world workloads. Our evaluation demonstrates that per-bank regulation effectively mitigates adversarial bank contention and achieves a 5.74$\times$ average throughput improvement for best-effort workloads over traditional bank-oblivious approaches while providing the same-level of performance isolation guarantees for real-time workloads.

\end{abstract}

\pagestyle{plain}
\section{Introduction} \label{sec:intro}

Modern multicore real-time systems face a fundamental challenge: unregulated sharing of DRAM can cause severe and unpredictable interference among applications. When multiple cores concurrently access memory, their requests contend for limited DRAM resources, creating interference that can significantly degrade application performance and undermine predictable timing guarantees. 

A substantial body of prior work has sought to mitigate memory interference through bandwidth regulation. State-of-the-art solutions~\cite{yun2013memguard,zuepke2023mempol,izhbirdeev2024memcore} aim to provide isolation guarantees by enforcing per-core bandwidth budgets at varying levels of granularity. These approaches have been effective in reducing interference, but they treat DRAM as a monolithic resource, ignoring its inherently parallel, banked organization. This structure-unaware abstraction leads to overly pessimistic performance because it regulates bandwidth globally rather than at the actual points of contention.

Our work is motivated by a simple but critical observation: DRAM interference is fundamentally a bank-level phenomenon. Requests to different banks can proceed largely in parallel, whereas requests targeting the same bank must be serialized. As a result, contention within a single bank dominates worst-case interference.

This observation suggests a fundamental shift in regulation strategy: rather than controlling aggregate DRAM bandwidth, regulation should operate at the granularity of individual banks. Aligning regulation with DRAM’s internal structure enables strong temporal isolation while preserving parallelism and improving overall system throughput.

In this paper, we analytically and empirically characterize the worst-case, guaranteed memory bandwidth~\cite{yun2015memory} of several generations of DRAM on representative commercial off-the-shelf multicore platforms. 
We show that this guaranteed bandwidth of modern DRAM is surprisingly small---and effectively constant across generations and  configurations---because it is dictated by the worst-case bandwidth of a single DRAM bank.

Building on this insight, we construct a single-bank memory performance attack that induces severe cross-core slowdowns across hardware platforms, even though the attacker generates minimal aggregate traffic. This result exposes a fundamental weakness in existing bandwidth-regulation approaches and demonstrates the need for new regulation approaches aligned with DRAM’s internal bank structure.

We propose a per-bank memory bandwidth regulation approach that enables predictable and high-performance use of shared DRAM in real-time systems. Our approach applies regulation at the bank level, scaling the throughput of best-effort tasks with the number of banks while preserving the temporal isolation required by real-time workloads.

We implement a novel per-bank bandwidth regulator design within an open-source RISC-V SoC framework~\cite{amid2020chipyard} and evaluate it on a Xilinx UltraScale+ VCU118 FPGA~\cite{vcu118} using FireSim~\cite{karandikar2018firesim}. We study the effects of bank-level contention attacks and evaluate the regulator’s ability to protect real-time tasks from such interference. Our results show that per-bank regulation effectively neutralizes bank contention attacks while providing, on average, a 5.74$\times$ throughput improvement for best-effort tasks over an all-bank regulation approach.

In summary, we make the following contributions:

\begin{itemize}
\item We characterize and empirically evaluate the guaranteed memory bandwidth of multiple DRAM technologies on representative multicore platforms. This includes reverse-engineering sophisticated XOR-based address mapping schemes and developing a precise methodology for measuring guaranteed bandwidth.
\item We demonstrate the limitations of existing memory-bandwidth regulation solutions through a single-bank memory performance attack and evaluate its impact on real hardware through extensive empirical analysis.
\item We design and implement a per-bank memory bandwidth regulator within a RISC-V SoC, evaluate it using a cycle-accurate, FPGA-accelerated full-system simulator, and show its ability to provide strong isolation and significant performance improvements. 
\item We release our tools, benchmarks, evaluation scripts, and proposed per-bank DRAM regulator hardware design as open source.\footnote{\url{https://github.com/CSL-KU/per-bank-dram-bru}}
\end{itemize}

The remainder of this paper is organized as follows. Section~\ref{sec:background} provides background information. Section~\ref{sec:guaranteed} characterizes guaranteed memory bandwidth. Section~\ref{sec:threatmodel} presents and evaluates a memory performance attack. Section~\ref{sec:solution} introduces our regulator design, with implementation details in Section~\ref{sec:implementation} and evaluation results in Section~\ref{sec:eval}. Section~\ref{sec:discussion} discusses broader implications. Section~\ref{sec:related} reviews related work, and Section~\ref{sec:conclude} concludes the paper.

\section{Background} \label{sec:background}
In this section we discuss modern DRAM structure, DRAM bandwidth regulation, and guaranteed memory bandwidth.

\subsection{DRAM Structure} \label{sec:dram-struct}

Modern DRAM systems are organized hierarchically, in channels, ranks, and banks, and achieve high performance via parallelism.
A primary driver of parallelism is the DRAM bank---a 2D memory array composed of rows and columns~\cite{memorysystems}.

Memory accesses to a DRAM bank proceed through a sequence of three commands: ACTIVATE, CAS, and PRECHARGE. An ACTIVATE command opens a row by transferring its contents into the row buffer of the bank. A subsequent CAS command reads or writes the desired data from the row buffer.
Additional accesses to the same row can be serviced directly from the row buffer through repeated CAS commands. However, when an access targets a different row, 
a PRECHARGE command must first be issued to write the buffer contents back to the cell array before the next row can be activated. 
Because these commands are issued per bank, multiple banks can be accessed in parallel, allowing the DRAM subsystem to exploit significant bank-level parallelism under typical workloads.

Memory controllers receive memory requests from the CPU and other clients, and schedule them on DRAM while respecting numerous timing parameters defined by DRAM standards~\cite{jedec-ddr4,jedec-ddr5,jedec-lpddr4}. Modern memory controllers maintain separate read and write transaction queues and commonly use FR-FCFS scheduling~\cite{rixner2000memory} to maximize overall throughput. 
The controller also manages the read/write turnaround time of the shared bidirectional data bus. Switching the bus from write mode to read mode incurs the \textit{tWTR} (Write-to-Read) timing penalty. To amortize this cost, controllers issue writes in large batches, a behavior governed by high and low watermarking schemes~\cite{chatterjee2012turnaround}. When the number of outstanding writes exceeds the high watermark, the controller begins draining writes from the queue.

\subsection{Memory Bandwidth Regulation} \label{sec:regulation}
Memory bandwidth regulation is a well-established approach to tame the cross-core memory interference on multicore. Both software- and hardware-based solutions have been explored.
Software-based approaches typically rely on architectural performance counters and stalling cores when they exceed a memory-access budget; MemGuard~\cite{yun2013memguard} is the canonical example.

Hardware-based regulators, in contrast, implement bandwidth enforcement directly in hardware, enabling finer-grained control. 
Intel Memory Bandwidth Allocation (MBA)~\cite{intel-rdt} is a representative example, which has been widely deployed in recent Intel server processors.

Unfortunately, existing approaches treat the entire DRAM as a monolithic resource and overlook its internal structure and bank-level parallelism. As a result, they either provide weak worst-case guarantees or incur unnecessary inefficiencies and bandwidth waste.

In~\cite{sullivan2024perbankcache}, the authors characterize such structure-oblivious regulation schemes as \textit{all-bank} regulation and propose a structure-aware alternative, \textit{per-bank} regulation, showing its benefits for managing bandwidth within banked last-level caches (LLCs).

Since DRAM exposes even more parallelism than a banked LLC, extending per-bank regulation principles to main memory promises even greater improvements—an opportunity we explore in this work.

\subsection{Guaranteed Memory Bandwidth} \label{sec:bg_guaranteed}

While modern DRAM scaling has achieved impressive increases in peak bandwidth, these gains are primarily enabled by increasing the number of concurrently accessible banks. However, the worst-case bandwidth of an individual bank has remained largely stagnant.

Following the definition in~\cite{yun2015memory}, we call the bandwidth achievable in the worst case the \textit{guaranteed memory bandwidth}. The worst case arises when all requests access different rows within the same DRAM bank, repeatedly causing row misses. In such scenarios, two consecutive requests must be separated by $tRC$ (row-cycle time)~\cite{jedec-ddr4}.
Consequently, the guaranteed bandwidth $BW_g$ can be expressed as: 
\begin{equation}
\text{$BW_{g}=\frac{64}{tRC}$}
\label{eq:guaranteed-bw}
\end{equation}

where each request corresponds to one 64-byte cache line from the CPU. Note that the $tRC$ parameter is largely determined by the intrinsic electrical properties of the DRAM cell array (capacitance, resistance, and sensing delay). Therefore, its value has remained relatively constant---typically 45 to 65 ns---across successive memory generations~\cite{jedec-ddr4, jedec-ddr5, jedec-lpddr4, jedec-hbm}, even as interface speeds and parallelism have increased.

\section{Measuring Guaranteed Memory Bandwidth} \label{sec:guaranteed}

In this section, we present a DRAM bank-mapping reverse-engineering tool and an evaluation methodology that can accurately measure the guaranteed memory bandwidth.

To accurately measure the guaranteed memory bandwidth, one must generate successive memory requests to different rows within the same DRAM bank. In some recent works~\cite{pradhan2025predictable,zuepke2023mempol}, the authors attempted to measure the guaranteed bandwidth\footnote{In~\cite{pradhan2025predictable,zuepke2023mempol}, the term \emph{sustainable memory bandwidth} was used instead, but its definition is identical to that of the guaranteed memory bandwidth.} by increasing the step size of memory accesses in powers of two, to trigger consecutive row misses within the same DRAM bank. Unfortunately, this approach works only if the physical address bits are mapped to DRAM banks via a simple direct mapping (i.e., a physical address bit directly corresponds to a DRAM bank bit). When the DRAM controller employs sophisticated XOR-based address mapping schemes~\cite{zhang2000permutation}, as seen in many x86 and high-end ARM systems, which map DRAM banks by XORing multiple physical address bits, simply probing different step sizes fails to generate the desired worst-case memory access patterns, resulting in significant overestimations~\cite{pradhan2025predictable}.

\subsection{Reverse Engineering of DRAM Bank Map Functions}

Our DRAM bank-mapping reverse-engineering tool is a modified version of DRAMA~\cite{pessel2016drama}, an experimental framework for reverse-engineering the address-mapping functions used by DRAM controllers. It measures pairwise access latencies between different memory addresses to identify those that map to the same DRAM bank, which incur longer access times due to row-conflict penalties. Using this information, DRAMA reconstructs the mapping between physical address bits and DRAM banks through statistical and linear-algebraic analysis over the Galois Field with two elements~\cite{lidl1997finitefields}, or GF(2) for short. DRAMA was originally designed for x86 platforms, as it relies on x86-specific instructions---\texttt{CLFLUSH} and \texttt{RDTSC}---for cache management and timing.

To enable DRAMA on ARM architectures, we made several modifications. First, we replaced the x86 \texttt{CLFLUSH} instruction with ARM’s cache maintenance operation \texttt{DC CIVAC}, which invalidates cache lines by virtual address to the point of coherency so that memory accesses are served from DRAM rather than from caches. Second, ARM lacks a direct equivalent to x86’s \texttt{RDTSC} instruction, as its closest counterpart, reading from the virtual timer counter (\texttt{CNTVCT\_EL0}), does not provide the fine-grained resolution needed to distinguish row-conflict timings. To overcome this limitation, we employ a signal amplification technique that performs repeated accesses to the same address pairs and measures the aggregate access time. This amplification compensates for the timer’s coarser granularity while maintaining sufficient timing resolution to differentiate between row-buffer hits and conflicts.

In addition, we identified and fixed several major issues, including a logic error that prevented high-order address bits from being considered and a performance bottleneck caused by the GF(2) solver’s exponential time complexity, which previously made successful address recovery difficult. In contrast, our modified version is much faster, runs in polynomial time, and is more robust, enabling reliable and efficient map recovery. We call our modified version \textit{DRAMA++}\footnote{\url{https://github.com/CSL-KU/drama-pp}}.

\begin{table*}[htp]
\centering
\scriptsize

\begin{tabular}{|c|c|c|c|c|}
\hline

\textbf{Platform} & \textbf{Raspberry Pi 4} & \textbf{Raspberry Pi 5} & \textbf{Intel Coffee Lake} & \textbf{Jetson Orin AGX} \\ \hline
    \textbf{CPU}            & 4 $\times$ Cortex-A72     & 4 $\times$ Cortex-A76         & i7-8700 (6C/12T)           & 12 $\times$ Cortex-A78E       \\ \hline
\multirow{3}{*}{\textbf{DRAM}}   & 2GB LPDDR4-3200    & 4GB LPDDR4X-4267   & 4 $\times$ 8GB DDR4-2133        & 4 $\times$ 8GB LPDDR5-6400  \\ 
                        & 1-die $\times$ 8-banks   & 2-die $\times$ 8-bank    & 4 $\times$ 2-rank $\times$ 16-banks  & 4 $\times$ 4-die $\times$ 16-banks             \\ 
                        & = 8 banks  & = 16 banks & = 128 banks   & = 256 banks       \\ \hline
\textbf{Peak B/W (GB/s)} & 12.8   & 17.1          & 34.1             & 204.8                 \\ \hline
\textbf{tRC (ns)}       & 60                 & 60                 & 47                 &  60               \\ \hline
\textbf{Found Bank Map}  & 
\begin{tabular}[c]{@{}l@{}}b0:12\\ b1:13\\ b2:14\end{tabular} & 
\begin{tabular}[c]{@{}l@{}}b0:12\\ b1:13\\ b2:14\\ b3:31\end{tabular} & 
\begin{tabular}[c]{@{}l@{}}b0:7$\oplus$14 \\ b1:15$\oplus$20\\ b2:16$\oplus$21\\ b3:17$\oplus$22 \\ b4:18$\oplus$23 \\ b5:19$\oplus$24 \\ b6:8$\oplus$9$\oplus$12$\oplus$13$\oplus$18$\oplus$19 \\
\end{tabular} & 
\begin{tabular}[c]{@{}l@{}}b0:11$\oplus$14$\oplus$16$\oplus$20$\oplus$21$\oplus$22$\oplus$33\\ b1:9$\oplus$11$\oplus$12$\oplus$16$\oplus$19$\oplus$23$\oplus$27$\oplus$28\\ b2:12$\oplus$13$\oplus$18$\oplus$22$\oplus$25$\oplus$29$\oplus$30$\oplus$31\\ b3:10$\oplus$11$\oplus$12$\oplus$17$\oplus$19$\oplus$20$\oplus$23$\oplus$32\\ b4:10$\oplus$11$\oplus$13$\oplus$14$\oplus$18$\oplus$27$\oplus$28$\oplus$34\\ b5:11$\oplus$12$\oplus$13$\oplus$16$\oplus$19$\oplus$24$\oplus$33$\oplus$35\\ b6:10$\oplus$13$\oplus$7$\oplus$21$\oplus$24$\oplus$25$\oplus$26$\oplus$29$\oplus$34\\ b7:14$\oplus$15$\oplus$17$\oplus$21$\oplus$25$\oplus$28$\oplus$31$\oplus$34$\oplus$35\end{tabular} \\ \hline
\end{tabular}
\caption{Evaluated hardware platforms.}
\label{tab:platform_comparison}
\end{table*}

\subsection{Hardware Platforms and Found DRAM Bank Maps}

For our study, we used three ARM platforms—Raspberry Pi 4, Raspberry Pi 5, and Jetson Orin AGX—and one x86 platform, an Intel i7-8700 desktop.
Table~\ref{tab:platform_comparison} summarizes the key characteristics of these hardware platforms, including the DRAM bank address-mapping information obtained using our reverse-engineering tool. The DRAM organization and the $tRC$ timing parameters of the first three platforms were obtained from the datasheets of the actual memory modules that we confirmed were used in our systems. For the Jetson Orin AGX, however, these parameters are our best estimates, as we could not find publicly available information and the memory chips were not visible without disassembling the entire system.
Note that all platforms employ different memory technologies—LPDDR4, LPDDR4X, DDR4, and LPDDR5—and the number of DRAM banks also varies significantly, from 8 to 256, representing a wide spectrum of systems.

The Raspberry Pi 4 and Raspberry Pi 5 platforms employ simple direct mapping schemes that use a subset of physical address bits to directly map DRAM banks.
On the Raspberry Pi 4, physical address bits 12, 13, and 14 are mapped to three DRAM bank address bits in the memory controller, corresponding to the eight banks of the LPDDR4 memory chip~\cite{pi4_mt53d512m32d2ds}.
The Raspberry Pi 5\footnote{Raspberry Pi 5’s bootloader supports four different DRAM bank maps~\cite{raspberry_document}. Recently, the default map was changed to map higher-order bits to banks
~\cite{raspberry_pi5_numa}. 
Our mapping corresponds to the older default map function.} uses physical address bits 12, 13, 14, and 31 to map its sixteen banks of its dual-die LPDDR4X memory~\cite{pi5_k4ube3d4ab-mgcl}.

In contrast, the Intel i7-8700 desktop and the Jetson Orin AGX employ more complex XOR-based address-mapping schemes. In the Intel desktop platform, four 8 GB dual-rank DDR4 DIMMs are installed, resulting in a total of 128 banks that are mapped using seven XOR-based mapping functions derived from multiple physical address bits. The Jetson Orin AGX platform, with a total of 256 banks, uses even more complex XOR-based mapping functions. 

\subsection{Bank-Aware Parallel Linked-List (PLL) Benchmark}

We use a modified version of the Parallel Linked-List (PLL) benchmark~\cite{bechtel2021memory} to measure the guaranteed bandwidth. PLL traverses multiple independent linked lists concurrently, where pointer-chasing dependencies within each list ensure that only one outstanding memory request exists per list.

Our modification in this work adds support for XOR-based address-mapping functions 
when creating linked-list entries, allowing them to be allocated on specific user-controlled DRAM banks.

\begin{algorithm}
\caption{Physical Address to Bank Conversion}
\label{alg:paddr_to_bank}
\begin{algorithmic}[1]
\Function{paddr\_to\_bank}{$paddr$}
    \State $bank \gets 0$
    \For{$i = 0$ \textbf{to} $|functions|-1$}
        \State $res \gets 0$
        \For{\textbf{each} $bit\_pos$ \textbf{in} $functions[i]$}
            \State $res \gets res \oplus ((paddr \gg bit\_pos) \& 1)$
        \EndFor
        \If{$res = 1$}
            \State $bank \gets bank \;|\; (1 \ll i)$
        \EndIf
    \EndFor
    \State \Return $bank$
\EndFunction
\end{algorithmic}
\end{algorithm}

Algorithm~\ref{alg:paddr_to_bank} shows the physical address to DRAM bank conversion function in our modified PLL benchmark. Here, $functions[i]$ refers to 
the XORed physical address bits that determine the $i$-th DRAM bank bit ($b_i$ in  Table~\ref{tab:platform_comparison}). 
For example, AGX’s $functions[0] = \{11,14,...,33\}$.

By varying the number of lists, the PLL benchmark explores the degree of exploitable memory-level parallelism (MLP) available in the memory hierarchy. This parallelism is influenced by the core’s ability to generate concurrent requests and the inherent parallelism of the DRAM banks.

However, when the number of DRAM banks available for access is constrained,
the exploitable MLP is likewise limited, which in turn affects performance. If it is restricted to a single bank, all parallel requests from the core are directed to that bank, creating a severe bottleneck. Moreover, because the addresses are randomly shuffled, they are likely to target different rows within the bank, producing the worst-case access pattern required to measure the guaranteed bandwidth, as defined in Section~\ref{sec:bg_guaranteed}.

\begin{figure*}[htp] 
  \centering
  \begin{subfigure}{0.49\textwidth}
    \includegraphics[width=\textwidth]{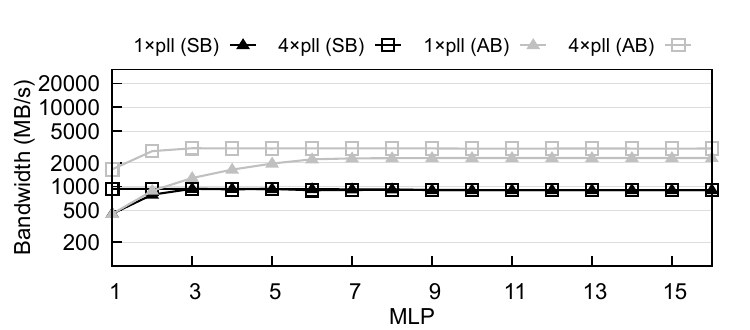}
    \caption{ Raspberry Pi 4 }
    \label{fig:pi4-mlp}
  \end{subfigure}
  \hfill
  \begin{subfigure}{0.49\textwidth}
    \includegraphics[width=\textwidth]{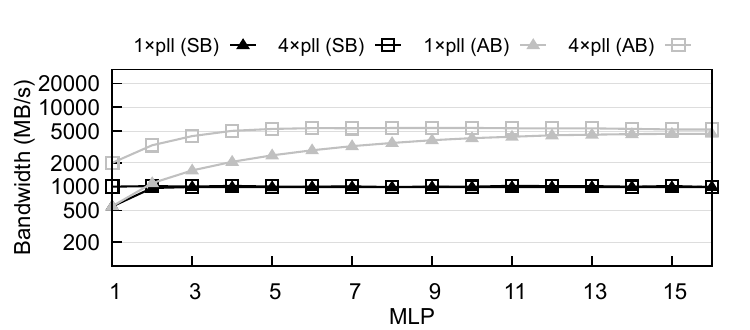}
    \caption{ Raspberry Pi 5 }
    \label{fig:pi5-mlp}
  \end{subfigure}

  \vspace{0.1cm}
  \begin{subfigure}{0.49\textwidth}
    \includegraphics[width=\textwidth]{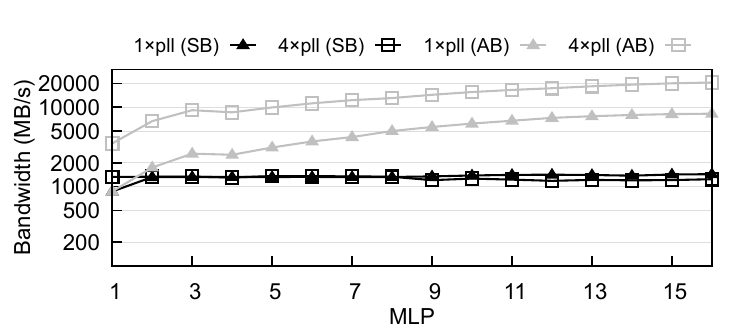}
    \caption{ Intel Coffee Lake 
    }
    \label{fig:intel-mlp}
  \end{subfigure}
  \hfill
  \begin{subfigure}{0.49\textwidth}
    \includegraphics[width=\textwidth]{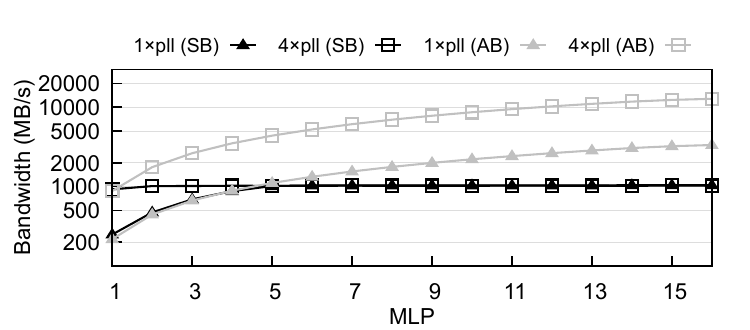}
    \caption{ Jetson Orin AGX }
    \label{fig:orin-agx-mlp}
  \end{subfigure}

  \caption{Measured bandwidth (logarithmic scale) of \emph{PLL} benchmarks as a function of MLP
  }
  \label{fig:all-mlptest}
\end{figure*}

\subsection{Evaluation Results}

Using the modified PLL benchmark, we conduct the following experiments to measure the guaranteed bandwidth of each hardware platform. For each platform, we swept the MLP (i.e., the number of lists $L$) from 1 to 16 and measured the aggregate bandwidth under four different configurations: 

\begin{itemize}
    \item \textit{1$\times$pll (SB)}: 1 instance targeting 1 (single) bank
    \item \textit{4$\times$pll (SB)}: 4 instances each targeting the same 1 bank
    \item \textit{1$\times$pll (AB)}: 1 instance targeting all banks
    \item \textit{4$\times$pll (AB)}: 4 instances each targeting all banks
\end{itemize}

Figure~\ref{fig:all-mlptest} shows the results. Note that X-axis represents the number of linked-lists $L$, which determines the MLP of the PLL benchmark. Y-axis is the measured bandwidth reported by the PLL. For 4 instance results, we report the aggregate bandwidth. In all experiments, each PLL instance was pinned to a dedicated core.

Let us first consider \textit{1$\times$pll (AB)} where the PLL benchmark can access all available DRAM banks on the platform. In this configuration, as the MLP increases, the measured bandwidth also increases---up to a point. The increase in bandwidth represents the core's (or LLC's) ability to extract memory-level parallelism as the MLP is not constrained by the parallelism of the DRAM banks. In \textit{1$\times$pll (SB)}, however, the bandwidth quickly saturates at around 1 GB/s. This is because a single DRAM bank cannot provide sufficient parallelism even though the core generates parallel requests. 

Similarly, in \textit{4$\times$pll (AB)}, the aggregate bandwidth further improves before saturation. This means that the multiple cores can extract additional parallelism from the memory hierarchy. 
In \textit{4$\times$pll (SB)}, however, adding more instances (i.e., more cores) does not increase aggregate bandwidth at all (Note that both (SB) lines are virtually indistinguishable except at MLP 1). This is because the maximum bandwidth is constrained by a single DRAM bank's maximum---i.e., the guaranteed bandwidth. 

\begin{table}[h!]
\centering
\begin{tabular}{|c|c|c|c|c|}
\hline
\textbf{Platform} & \textbf{Pi 4} & \textbf{Pi 5} & \textbf{Intel} & \textbf{AGX} \\ \hline
\textbf{Theory} & 1067 & 1067 & 1362 & 1067 \\ \hline
\textbf{Measured } & 939 & 945 & 1324 & 1042 \\ \hline
\end{tabular}
\caption{Guaranteed bandwidth (MB/s)}
\label{tab:bandwidth_guaranteed}
\end{table}

Table~\ref{tab:bandwidth_guaranteed} compares the theoretical guaranteed bandwidth (calculated using Eq.~\ref{eq:guaranteed-bw}) with our measured results—that is, the maximum observed bandwidth in either 1$\times$pll (SB) or 4$\times$pll (SB). Note that both the calculated and measured values are closely aligned, at around 1 GB/s on ARM platforms using LPDDR memories and 1.3 GB/s on the Intel platform with DDR4 memory. This close alignment supports the validity of our measurement method.

In summary, we have demonstrated that despite the high degree of parallelism and high peak bandwidth in modern memory systems, the guaranteed memory bandwidth remains small and stable, as it is determined by the worst-case bandwidth of a single DRAM bank. In the following, we will explore how this observation can be exploited to induce severe memory performance interference in multicore systems.

\section{Single-Bank Memory Performance Attack} \label{sec:threatmodel}

In this section, we investigate how the execution time of an application can be affected by interfering memory requests from co-running applications at the DRAM bank level.

We begin by posing the following question: \textit{What is the most effective way to delay an application accessing DRAM?}

Suppose the application under study, which we call the \textit{victim}, generates eight concurrent memory requests to an eight-bank DRAM. Without interference, the victim's requests can ideally be processed in parallel—one per bank---taking one unit of time in total. Now suppose that an interfering application, which we call the \textit{attacker}, also generates eight parallel requests simultaneously, potentially interfering with the victim's requests. If the attacker's requests are evenly distributed across the eight banks, the maximum delay these interfering requests can cause is bounded by the delay at a single bank---one unit time---rather than the aggregate of all banks. On the other hand, if the attacker's requests are directed to a single bank, then the worst-case delay the victim may experience is eight, as the victim's single request to that bank might be delayed by up to eight time units. In this case, even though the victim's other requests experience no delay, they have no impact on the overall execution time, which is determined by the single bank suffering the maximum delay.

From this hypothetical thought experiment, we therefore expect that the worst-case delay will occur when interfering memory requests are concentrated on a single DRAM bank---a hypothesis that we will validate next.

\subsection{Methodology}

We evaluate the impact of bank-level interference using one victim process running on a dedicated core and three attacker processes running concurrently on separate cores. The attackers execute the PLL benchmark that either accesses all DRAM banks (AB) or a single bank (SB) and issues either read or write requests. We denote these configurations as \textit{ABr} (all-bank read), \textit{ABw} (all-bank write), \textit{SBr} (single-bank read), and \textit{SBw} (single-bank write), respectively.

For each experiment, we first measure the victim's execution time in isolation. We then launch three attacker processes on dedicated cores (cores 1–3) and re-run the victim on core 0 to measure its execution time under interference. We report the victim's slowdown ratio together with the aggregate bandwidth consumed by the attackers.\footnote{Throughout the paper, all experiments were repeated multiple times, but we report only one data point for each as no significant run-to-run variation was observed. Our code repository includes scripts for reproduction study.}

\subsection{Effects on Synthetic Victim}~\label{sec:attack_synthetic}

\begin{figure*}[htp]
  \centering
  \begin{subfigure}{0.49\textwidth}
    \includegraphics[width=\textwidth]{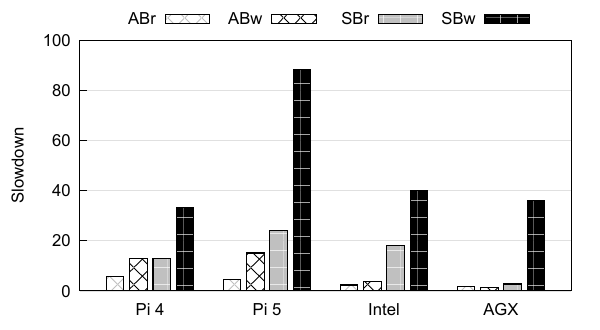}
    \caption{ Victim slowdown. }
    \label{fig:platform-bw}
  \end{subfigure}
  \begin{subfigure}{0.49\textwidth}
    \includegraphics[width=\textwidth]{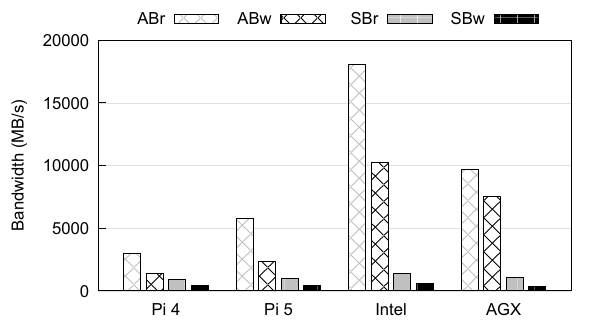}
    \caption{Attacker bandwidth. }
    \label{fig:platform-sd}
  \end{subfigure}
  \caption{Victim (\textit{Bandwidth}) slowdown and attacker bandwidth consumption of all-bank (AB) and single-bank (SB) attacks.}
  \label{fig:dram-bank-contention}
\end{figure*}

In this experiment, we use the \emph{Bandwidth} benchmark from IsolBench~\cite{valsan2016taming} as the victim. The benchmark simply performs sequential reads over a large array---twice the size of the LLC---forcing most of its accesses to require DRAM accesses.

Figure~\ref{fig:dram-bank-contention} shows the results on all four platforms we tested. Note first that, compared to the all-bank attacks (ABr and ABw), the single-bank attacks (SBr and SBw) are significantly more effective in delaying the victim, as expected.
This is despite the fact that the single-bank attacks generate significantly \emph{less} bandwidth.
For example, on the Pi 5, SBw attackers cause the highest slowdown of the victim---88$\times$---while generating the least amount of memory bandwidth, less than 440 MB/s.\footnote{Write bandwidth observed from the application. It does not account for write-induced cache-line refill read traffic. The combined read and write traffic at the DRAM is still upper-bounded by the guaranteed bandwidth.} In fact, among the four attackers we tested, the least effective one (ABr) consumes the highest memory bandwidth, while the most effective one (SBw) consumes the least---exactly the \textit{opposite} of conventional wisdom.

\begin{figure*}[htp]
  \centering

  \begin{subfigure}{0.49\textwidth}
    \includegraphics[width=\textwidth]{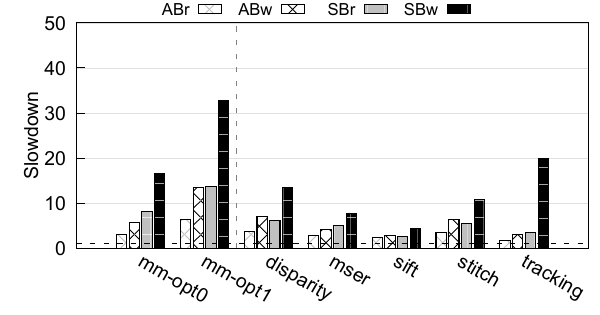}
    \caption{ Raspberry Pi 4 }
    \label{fig:pi4-slowdown}
  \end{subfigure}
  \hfill
  \begin{subfigure}{0.49\textwidth}
    \includegraphics[width=\textwidth]{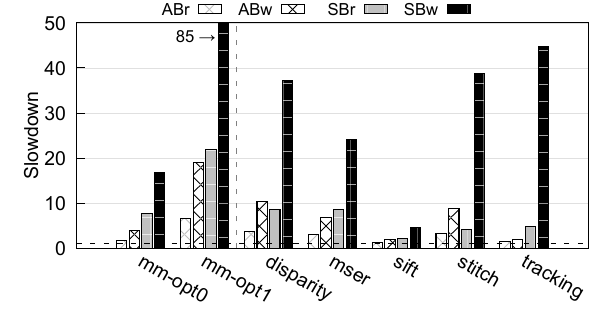}
    \caption{ Raspberry Pi 5 }
    \label{fig:pi5-slowdown}
  \end{subfigure}

  \vspace{0.3cm}
  \begin{subfigure}{0.49\textwidth}
    \includegraphics[width=\textwidth]{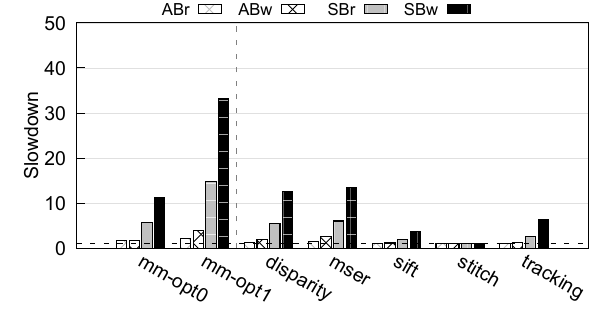}
    \caption{ Intel Coffee Lake
    }
    \label{fig:intel-slowdown}
  \end{subfigure}
  \hfill
  \begin{subfigure}{0.49\textwidth}
    \includegraphics[width=\textwidth]{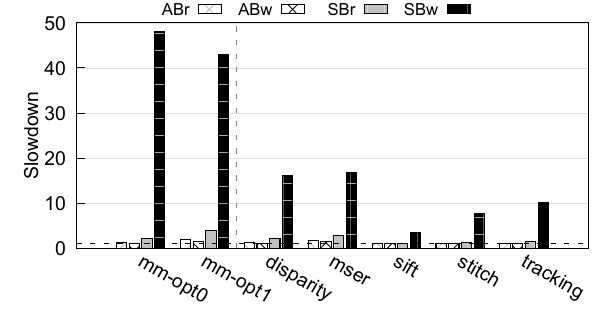}
    \caption{ Jetson Orin AGX }
    \label{fig:orin-agx-slowdown}
  \end{subfigure}

  \caption{Victim (X-axis) slowdown caused by all-bank (AB) vs. single-bank (SB) attacks.
  }
  \label{fig:all-slowdowns}
\end{figure*}

\subsection{Effects on Real-world Applications}~\label{sec:attack_real}

In this experiment,
we use the same experimental setup, but use different victims from a set of real-world benchmarks: two matrix multiplication kernels~\cite{matmult-githubrepo}
and five workloads from SD-VBS~\cite{sdvbs} (with \textit{fullhd} input).
The two matrix multiplication kernels differ in their access pattern. The \textit{mm-opt0} is a na\"ive implementation,
which suffers from poor locality.
In contrast, \textit{mm-opt1} is an optimized version
with improved spatial locality along the innermost loop.
The SD-VBS benchmarks represent
diverse memory access patterns.

Figure~\ref{fig:all-slowdowns} shows the results.
Across all platforms and benchmarks, SB attacks—especially \textit{SBw}—consistently cause extremely high victim slowdowns despite consuming the least memory bandwidth (we omit the attacker-bandwidth figure here because it is similar to Figure~\ref{fig:platform-sd}). Although the degree of effectiveness varies across platforms due to microarchitectural differences and across benchmarks due to differences in memory-access characteristics, the overall trends are consistent with those observed for the synthetic victim in the previous experiment (Section~\ref{sec:attack_synthetic}), further supporting our hypothesis.

These results present a major challenge for system designers seeking to provide real-time guarantees using existing memory bandwidth regulation techniques, which limit (throttle) memory bandwidth without distinguishing how that bandwidth is distributed across DRAM banks. To provide worst-case performance guarantees, one must assume the worst-case scenario in which traffic is concentrated on a single bank. However, doing so requires the bandwidth budget to be extremely low—below the guaranteed bandwidth—thereby wasting much of the available memory bandwidth in typical cases where traffic is more evenly distributed across banks. We argue that this limitation arises fundamentally because all existing bandwidth regulation techniques are \textit{bank-oblivious}, which we aim to address in the following.
\section{Per-Bank DRAM Bandwidth Regulation} 
\label{sec:solution}

As discussed earlier, all state-of-the-art memory bandwidth regulators operate in a bank-oblivious manner, which we call the \textit{all-bank} approach. This approach treats memory as a monolithic structure, ignoring the parallel, banked organization of modern main memory.

In contrast, we propose a \textit{per-bank} approach, which regulates accesses to each DRAM bank individually. In the worst-case scenario—i.e., when traffic is concentrated on a single bank—the per-bank regulation approach can provide the same isolation guarantees as the all-bank regulation (given the same bandwidth budget). At the same time, it greatly improves overall memory throughput in typical cases where traffic is more evenly distributed across banks, since the same budget, $B_{\text{per-bank}}$, applies to each individual bank. As a result, the maximum usable bandwidth, $BW_{\text{max}}$, scales linearly with the number of banks $N_{\text{bank}}$ as follows:
\begin{equation}
BW_{\text{max}} = B_{\text{per-bank}} \times N_{\text{bank}}
\label{eq:per-bank-scaling}
\end{equation}

In the following, we present the design of our proposed per-bank DRAM bandwidth regulator. 

\subsection{Regulator Placement}
We first start by discussing where such a regulator could be placed. 
Ideally, DRAM bandwidth should be controlled by regulating the rate at which LLC misses are served to memory---between the LLC and the memory controller.

A simple option is a drop-in module, similar to BRU~\cite{farshchi2020bru,sullivan2024perbankcache} for cache bandwidth regulation, placed on the interconnect between the LLC and memory controller. However, we find this to be suboptimal for DRAM regulation since the regulator's clients are the shared LLC cache banks, not the cores. Stalling an entire LLC bank due to one core exceeding its budget would block unrelated requests from other cores, severely degrading performance.

Adding request queues between the LLC and memory controller can potentially address such unnecessary stalls, but this adds significant area overhead, especially when regulating both reads and writes.

To avoid these issues, we place our regulator inside the shared LLC, where it controls when Miss Status Holding Registers (MSHRs) are allowed to issue requests to memory. This enables per-core, per-bank bandwidth regulation without requiring additional queuing structures.

\subsection{Regulation Scheme}
Our regulator uses a fixed-rate regulation scheme similar to MemGuard~\cite{yun2013memguard}. This means that the regulator operates with a fixed regulation period (in cycles) and a per-period access budget. Regulation works by counting the number of memory accesses issued by a core within the current period. If that count exceeds the assigned access budget, all subsequent requests from the core are prevented from reaching memory. The counters are reset at the beginning of each new period, effectively replenishing the budgets and allowing cache misses to proceed as usual. The per-bank bandwidth budget, $B_{\text{per-bank}}$, can be calculated as follows: 

\begin{equation}
B_{\text{per-bank}} = \frac{N_{\text{acc}}}{P} \times G \times f
\label{eq:per-bank-budget}
\end{equation}

where $N_{\text{acc}}$ is the number of accesses per period $P$ (in cycles), $G$ is the access granularity (in bytes, typically a 64-byte cache line), and $f$ is the frequency of the clock domain in which the regulator resides, which in our current design is the frequency of the LLC.
The choice of the fixed-rate regulation scheme also informs the rest of our design.

\subsection{Regulator Design}
Our design adopts regulation domains, as in ~\cite{farshchi2020bru}, which enables arbitrary grouping of cores to be regulated together as a group. 
Therefore, each regulation domain has its own unique budget configuration register. The number of domains can be configured at design synthesis.

Our regulator requires a way to attribute accesses to cores. For this, our design includes a ``tagging unit" that sits immediately after the cores. This unit exposes an MMIO interface to configure which cores are in which regulation domain. The tagging unit adds the domain information to each request that is sent from the cores to the shared LLC. With this domain information, the regulation unit can attribute LLC misses to the correct counters and selectively enable or disable regulation for a given domain. 

For regulation configuration, we expose a set of MMIO control registers.
A dedicated register is used to set the global regulation period $P$, while another set of registers configures the access budget, $N_{\text{acc}}$, for each domain during the period.

\section{Implementation}
\label{sec:implementation}
\begin{figure}[htp]
    \centering
    \includegraphics[width=0.48\textwidth]{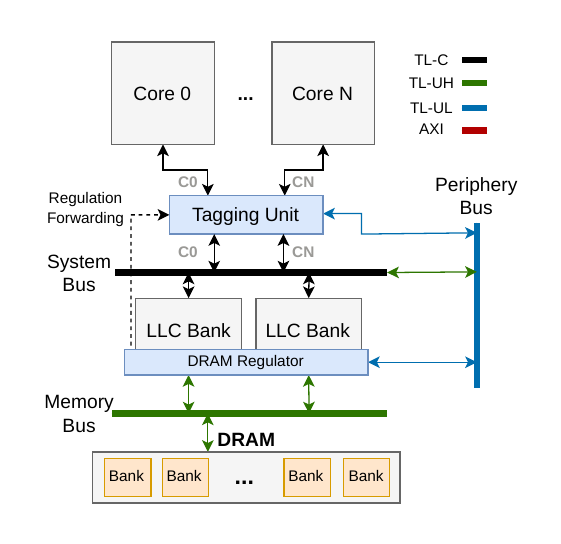}
    \caption{Rocket SoC with our DRAM bandwidth regulator}
    \label{fig:design}
\end{figure}

We implement our design in a RISC-V Rocket Chip SoC~\cite{krste2016rocket} using the Chipyard~\cite{amid2020chipyard} framework. Figure~\ref{fig:design} offers a high-level view of our design integrated into this SoC context. This section discusses the implementation details of our work.

\subsection{Cache Integration}
The LLC available in Chipyard is the \textit{Rocket Chip inclusive cache}~\cite{inclusive-L2}. As the name suggests, it is an inclusive cache with a configurable number of banks and serves as the shared LLC in our simulated SoC. Supporting directory-based coherence, this cache is implemented as a hierarchical set of modules, with a top-level module wrapping the cache banks, and each cache bank containing its own scheduler, MSHRs, directory, and related components.

In our design, the top-level module of the LLC contains the MMIO interface and counters for regulation. We count TileLink \textit{AcquireBlock} requests (reads) sent to main memory and attribute them to the appropriate domain and bank counters. When a bank access counter exceeds the configured budget for a domain, a throttle signal is asserted and propagated to the cache schedulers. Within the schedulers, this throttle signal is used by the round-robin arbiter that selects which MSHR will service its request next. By default, the arbiter considers an MSHR schedulable if all the cache resources it requires are available. We modify this scheduling decision to also gate schedulability based on the throttle signals.

\subsection{Regulation Forwarding}
Stalling MSHRs enables precise control over the bandwidth that a core can send to the shared main memory. This mechanism alone, however, is insufficient, as residual contention can still occur within the cache banks themselves. To address this issue, we enable regulation of cache bandwidth via throttle signals forwarded from the DRAM regulator up to the tagging unit (see Figure~\ref{fig:design}). With these forwarded signals, the tagging unit can control cache bandwidth
by stalling \textit{AcquireBlock} requests from cores in a throttled domain when the corresponding throttle signal is asserted. This set of signals consists of $D \times N_{\text{bank}}$ bits, where $D$ is the number of domains and $N_{\text{bank}}$ is the number of DRAM banks.
\section{Evaluation} \label{sec:eval}

In this section, we evaluate the proposed per-bank regulator in terms of its worst-case guarantees and average throughput.
\subsection{Experimental Setup}

We use FireSim~\cite{karandikar2018firesim} as our simulation environment. FireSim is an FPGA-accelerated, cycle-exact, full-system simulator. It enables simulation of a complete SoC, with the FPGA operating at 100 MHz while the simulated target runs at a configurable frequency. Simulation results are not affected by the FPGA speed limitation, as a specialized compiler infrastructure decouples the target design from the FPGA host~\cite{biancolin2019fased}, enabling realistic performance evaluation.

\begin{table}[htpb]
	\centering
	\begin{tabular}{|c||p{5.5cm}|}
    	\hline
    	\multirow{2}{*}{Cores}  & 4$\times$BOOM, 1GHz, out-of-order, 2-wide, ROB: 64, LSQ: 16/16, L1: 16K(I)/16K(D), 6 mshrs \\
    	\hline
    	\multirow{2}{*}{Shared L2 Cache} & 1MB (16-way), 2 banks, 27 mshrs per bank, random replacement \\
    	\hline
    	\multirow{2}{*}{DRAM} & 4GB 1-rank 8-bank DDR3, FR-FCFS,  \\
                             &  Bank map: 9, 10, 11 (direct map); tRC = 47ns\\
    	\hline
	\end{tabular}
	\caption{Evaluation platform specifications.}
	\label{tab:fsim-specs}
\end{table}

Our simulation configuration is shown in Table~\ref{tab:fsim-specs}\footnote{We choose address mapping functions for LLC sets and DRAM banks to enable LLC set partitioning without DRAM bank co-partitioning. We also provision enough LLC MSHRs to avoid MSHR contention~\cite{valsan2016taming}.}. The SoC consists of four Berkeley Out-of-Order Machine (BOOM)~\cite{asanovi2015boom} cores in their medium (2-wide) configuration. We include the \textit{Rocket Chip inclusive cache}~\cite{inclusive-L2} in the design, which serves as a 1 MB shared last-level cache (LLC). The entire SoC is configured to operate at 1 GHz. For main memory, FireSim utilizes the FASED memory controller which uses a memory timing model to bridge the gap between FPGA and real DDR backing memory~\cite{biancolin2019fased}. We configure it to model a single-channel, single-rank DDR3 memory system with an FR-FCFS scheduling policy.
All simulations run on FireSim, running real RISC-V Linux kernel v6.2. We patch the kernel with PALLOC~\cite{yun2014palloc} to enable LLC set partitioning.

\subsection{FASED DRAM Controller Enhancements}
To effectively evaluate DRAM contention, it is important that the memory model in the simulation environment accurately represents real hardware. The baseline FASED memory controller~\cite{biancolin2019fased} implements a unified transaction queue, meaning that both read and write transactions are buffered together and sent to the command scheduler in FIFO order. While simple to implement, as discussed in Section~\ref{sec:dram-struct}, this design is unrealistic, as most modern memory controllers implement separate read and write buffers and watermark based batched scheduling~\cite{chatterjee2012turnaround} to improve performance.

We enhance the FASED memory controller by introducing separate transaction queues and write batching with parameterizable watermarks. To demonstrate the impact of these enhancements, we run a random-access, write-heavy workload (\textit{PLL}) on a simulated SoC. One run uses the baseline FASED controller, and the other uses our enhanced version. We then collect the number of bus mode switches that occur during the execution of the benchmark.

\begin{table}[htp]
\centering
\small
\begin{tabular}{l c}
\toprule
\textbf{FASED Version} & \textbf{Mode Switches} \\
\midrule
Baseline & 1,813,936 \\
Ours (w/ Batching) & 577,155 \\
\bottomrule
\end{tabular}
\caption{Write-batching impact on bus mode switches.}
\label{tab:dram-mode-switches}
\end{table}

Table~\ref{tab:dram-mode-switches} presents the results, showing a 3.14$\times$ reduction in mode switches when using the write-batching–enabled version of the model. Our enhancements enable more realistic memory performance evaluation, which we plan to release as open-source.

\subsection{Guaranteed Bandwidth in FireSim}
\label{sub:fsim-guranteed-bw}

We begin our evaluation by exploring the guaranteed DRAM bandwidth in our simulated SoC. Following Eq.~\ref{eq:guaranteed-bw}, we first calculate the theoretical guaranteed bandwidth of the simulated DRAM system.
For validation, we also run the same PLL workload on FireSim, targeting a single DRAM bank, as described in Section~\ref{sec:guaranteed}.

\begin{table}[h]
\centering
\small
\begin{tabular}{l c}
\toprule
Theory & 1362 \\
Measured & 1271 \\
\bottomrule
\end{tabular}
\caption{Guaranteed memory bandwidth (MB/s) in FireSim.}
\label{tab:fsim-guaranteed-bw}
\end{table}

Table~\ref{tab:fsim-guaranteed-bw} shows the results. As can be seen,
the measured bandwidth closely matches the expected theoretical number.
These results validate that the FASED memory model offers a realistic evaluation platform.

\subsection{Effects of DRAM Bank Contention}

In this experiment, we repeat the all-bank and single-bank attack experiment against a synthetic victim benchmark as described in Section~\ref{sec:attack_synthetic}---that is, we use \textit{Bandwidth} from IsolBench~\cite{valsan2016taming} as the \emph{victim} and use \textit{ABr} (all-bank read), \textit{ABw} (all-bank write), \textit{SBr} (single-bank read), and \textit{SBw} (single-bank write) as the attackers to understand the attackers' impact on the victim as well as the bandwidth usage of the attackers on our simulated RISC-V SoC platform.

\begin{figure}[htp]
  \centering
  \begin{subfigure}{0.4\textwidth}
    \includegraphics[width=\textwidth,trim={1.5cm 0 0.05cm 0},
  clip]{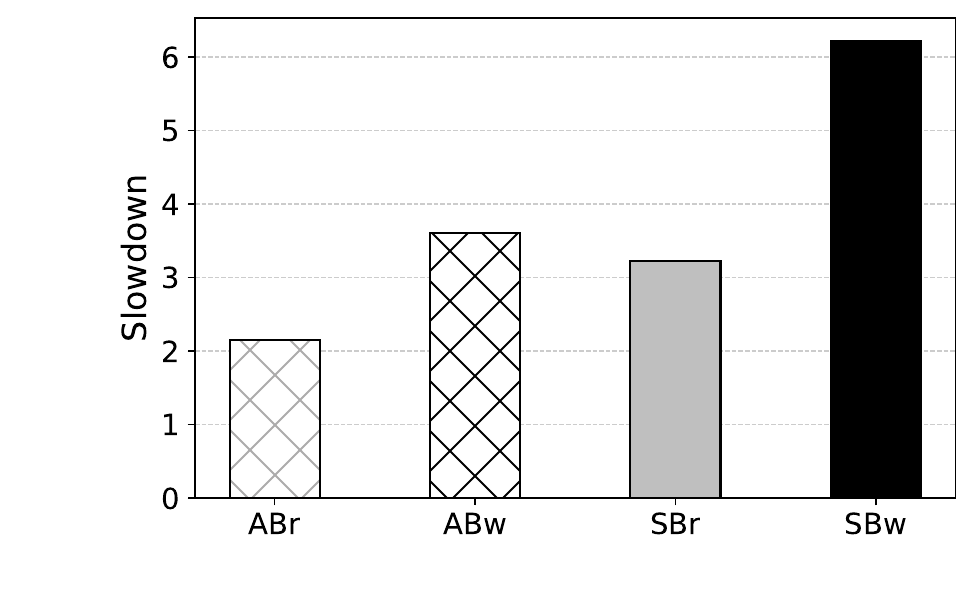}
    \caption{ Victim Slowdown. }
    \label{fig:fsim-victim-slowdown}
  \end{subfigure}
  \begin{subfigure}{0.4\textwidth}
    \includegraphics[width=\textwidth,trim={1.5cm 0 0.05cm 0},
  clip]{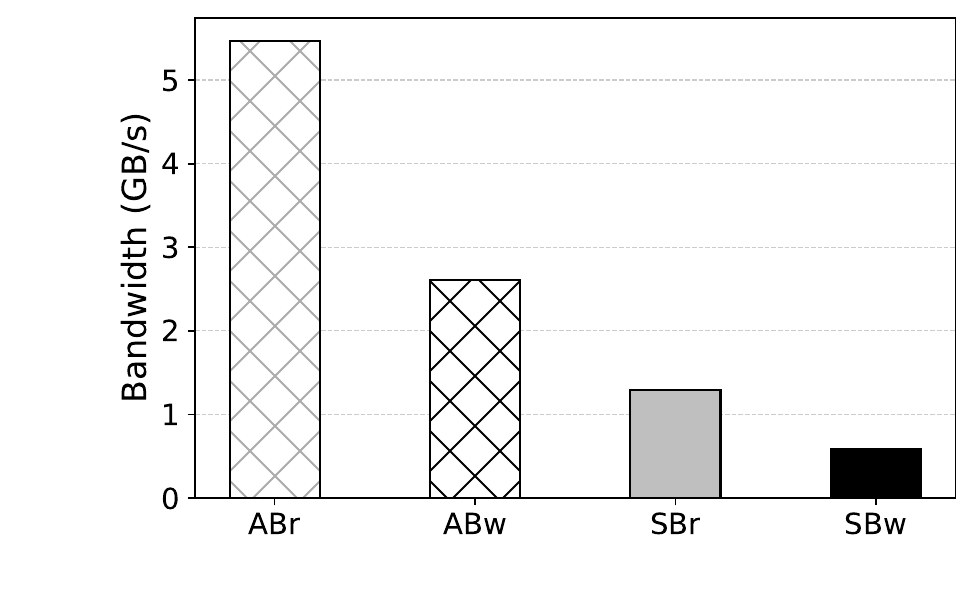}
    \caption{ Attacker Bandwidth. }
    \label{fig:fsim-attack-bandwidth}
  \end{subfigure}
  \caption{ Effects of all-bank (AB) and single-bank (SB) DRAM bank attacks against the synthetic victim on FireSim.}
  \label{fig:fsim-dram-bank-contention}
\end{figure}

Figure~\ref{fig:fsim-dram-bank-contention} shows the results. Note that the trends observed here closely mirror those from the real platform experiments (cf. Figure~\ref{fig:dram-bank-contention}). Specifically, \textit{ABr} causes the least slowdown ($\sim$2.1$\times$) to the victim while generating the highest memory bandwidth ($>$5 GB/s), whereas \textit{SBw} causes the greatest slowdown ($\sim$6.2$\times$) yet consumes the least bandwidth ($<$1 GB/s).

\subsection{Effects of Per-Bank Bandwidth Regulation}
\label{sub:fsim-per-bank-eval}

In this subsection, we evaluate the performance of the proposed \textit{per-bank regulator}.
For comparison, we also implement an \textit{all-bank regulator}, which is a modified version of our regulator where each domain maintains a single global access counter, instead of per-bank counters as in our proposed design. This effectively treats DRAM as a monolithic structure, similar to existing memory bandwidth regulation solutions.

In both regulators, we set the regulation period to \textit{1 ms} with a bandwidth budget of 53 MB/s, which we empirically (sweeping the budgets) determined to be the maximum budget needed to bound the worst-case victim slowdown to an acceptable level (less than 10\% of the solo execution time).
We configure the four cores into two regulation domains: \textit{real-time} and \textit{best-effort}.
The real-time domain includes only Core 0 and remains unregulated to ensure maximum real-time performance.
The best-effort domain includes the remaining three cores (Cores 1–3) and is regulated using the above settings to keep their interference to the real-time task within bounded limits. The LLC space is evenly partitioned between the real-time and best-effort domains, using PALLOC~\cite{yun2014palloc}, to avoid cache evictions between the two domains.

First, we examine the regulators' ability to provide isolation guarantees to the task in the real-time domain against interference from the all-bank and single-bank write attackers---\textit{ABw} and \textit{SBw}, respectively---on the best-effort domain. The rest of the experimental setup is identical to the previous one, except that the attackers are now regulated.

\begin{figure}[htp]
    \centering
    \includegraphics[width=0.4\textwidth]{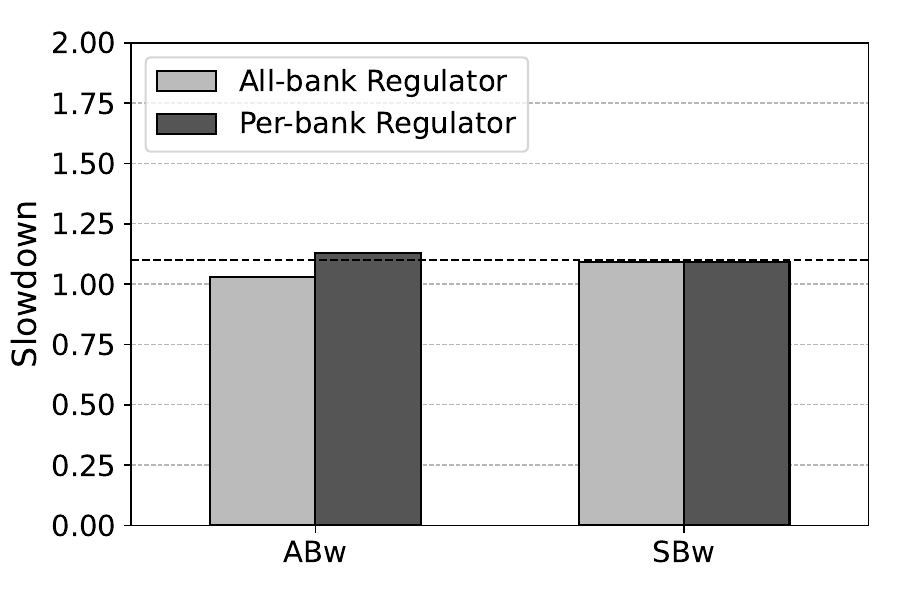}
    \caption{Victim slowdown under all-bank vs. per-bank regulation against \textit{ABw} and \textit{SBw} attackers.}
    \label{fig:victim-isol-1ms}
\end{figure}

Figure~\ref{fig:victim-isol-1ms} shows the results.
Both the all-bank and per-bank regulators limit the victim's slowdown to 1.1$\times$ the solo execution time against the single-bank write attack (\textit{SBw}).
Against the all-bank write attacker (\textit{ABw}), however, our per-bank regulator yields a slightly higher slowdown (1.13$\times$) than in the single-bank case, whereas the all-bank regulator limits the slowdown to 1.03$\times$. Relative to the 1.1$\times$ baseline, this means that the per-bank regulator incurs only a 3\% additional overhead due to delays outside the DRAM banks, while enabling the \textit{ABw} attacker—and any benign best-effort task—to achieve up to an 8$\times$ performance improvement over the all-bank regulator. We consider this a highly favorable tradeoff. We now turn to empirical results demonstrating the throughput gains of our per-bank regulator.

\begin{figure}[htp]
    \centering
    \includegraphics[width=0.45\textwidth]{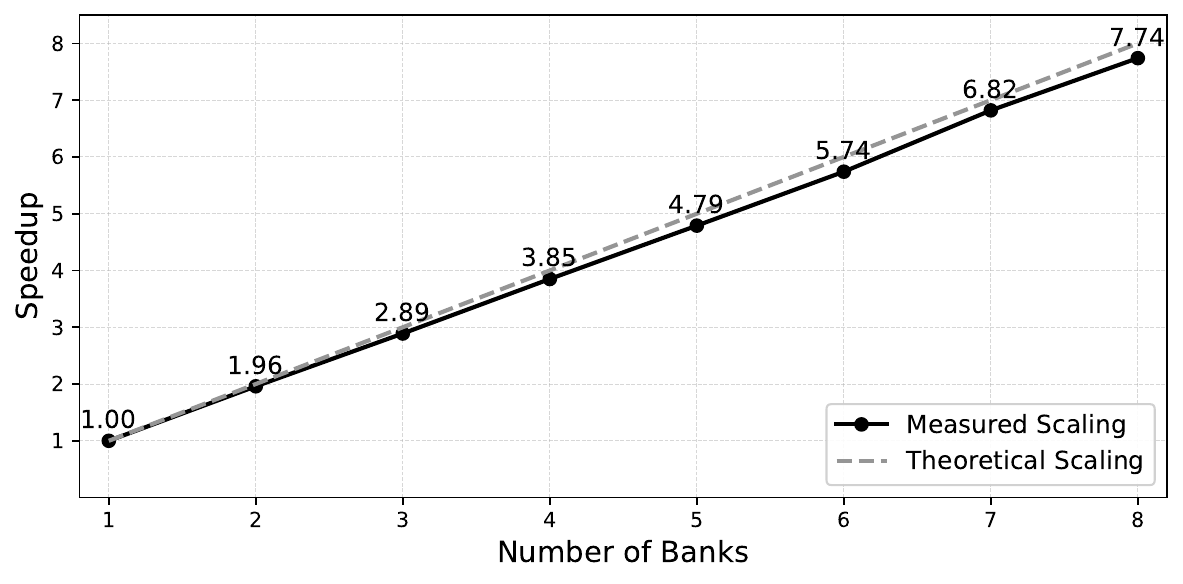}
    \caption{Per-bank regulation scaling.}
    \label{fig:per-bank-scaling}
\end{figure}

Figure~\ref{fig:per-bank-scaling} shows the speedup of the regulated attacker tasks on the best-effort domain as a function of the number of banks of the DDR3 memory.  The results clearly show that, as the number of DRAM banks increases, the observed speedup linearly increases, confirming that throughput scaling as predicted by Eq.~\ref{eq:per-bank-scaling}. Although it does not achieve the perfect linear scaling (at 8 banks, the speedup was 7.74$\times$) because accesses  to DRAM banks are not completely parallel, this shows the potential performance benefits of the per-bank approach.

The additional bandwidth provided by our per-bank regulation approach can be especially beneficial when real-world applications---unlike the specially engineered attackers used above---are executed on the best-effort domain.
To demonstrate this, we use a set of representative real-world workloads: two matrix multiplication kernels and five benchmarks from SD-VBS~\cite{sdvbs} (with \textit{fullhd} input), as described in Section~\ref{sec:attack_real}.
Each workload is executed on a single core in the \textit{best-effort} domain. We measure its runtime under both all-bank and per-bank regulation and compare the results against the unregulated baseline.

\begin{figure*}[htp]
    \centering
    \includegraphics[width=0.8\textwidth]{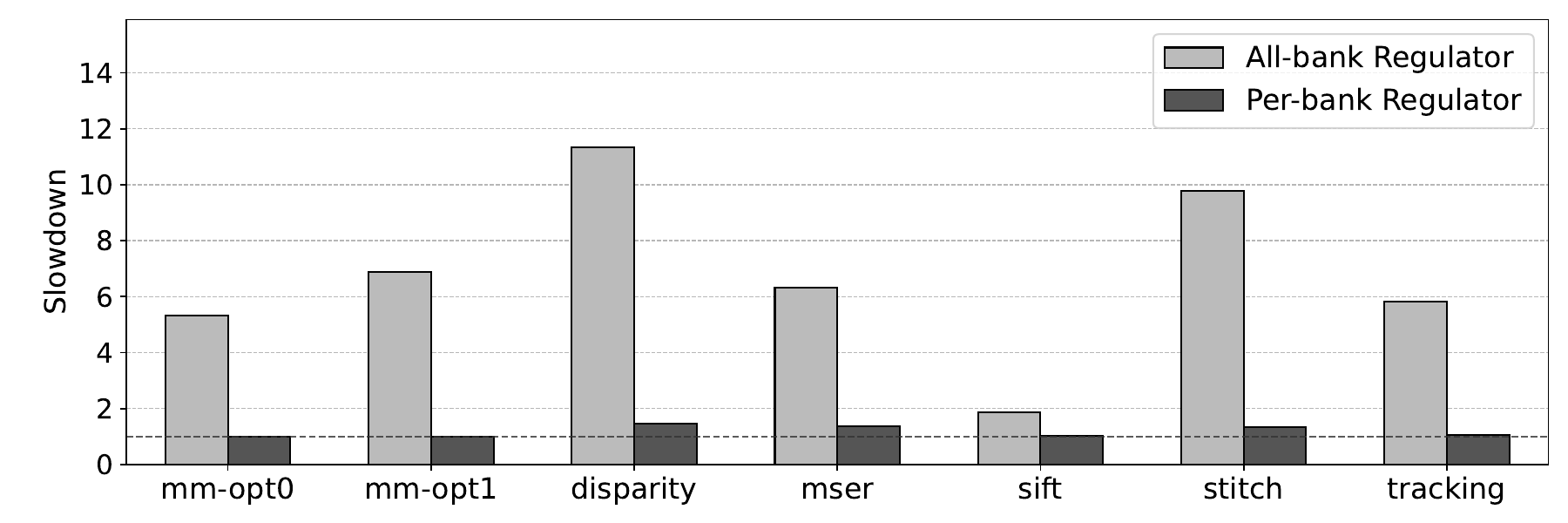}
    \caption{Average case (benign best-effort tasks) performance.}
    \label{fig:real-wrold-be-thruput}
\end{figure*}

Figure~\ref{fig:real-wrold-be-thruput} shows the results. As expected, our per-bank regulator achieves significant gains over the all-bank regulator across all benchmarks. Compute-intensive workloads such as \textit{sift} predictably show smaller improvements, whereas memory-intensive workloads such as \textit{disparity} show substantial gains. Overall, the per-bank regulator achieves an average performance improvement of 5.74$\times$ over the all-bank regulator.

In summary, we have shown that per-bank memory bandwidth regulation can provide isolation benefits comparable to those of all-bank regulation while offering substantially higher, linearly scalable memory performance benefits. As the parallelism in modern DRAM technologies continues to increase (for instance, HBM memories have thousands of DRAM banks), we believe the potential benefits of the per-bank approach will become even more pronounced.

\subsection{Implementation Overhead}
To quantify the implementation cost of our design, we synthesize a quad-core BOOM SoC with and without our regulator using Cadence Genus with Hammer~\cite{liew22hammer} automation, targeting the ASAP7 7nm technology node~\cite{clark16asap7}. In addition to the default 8-bank DRAM configuration, we also evaluate a 16 bank DRAM configuration to evaluate scalability.

\begin{table}[h]
\centering
\small
\begin{tabular}{c c c}
\toprule
\textbf{\# Banks} & \textbf{Area} & \textbf{Timing} \\
\midrule
8 & 0.35\% & 3\% \\
16 & 0.47\% & 3\% \\
\bottomrule
\end{tabular}
\caption{Area and timing overhead of per-bank regulator.}
\label{tab:area-timing}
\end{table}

Table~\ref{tab:area-timing} shows the results.
Compared to the baseline, our per-bank regulator incurs minimal area overhead (0.34-0.47\%) and modest timing overhead, reducing the maximum achievable clock frequency by 3\% across both bank configurations.
\section{Discussion} \label{sec:discussion}

In this section, we discuss how per-bank bandwidth regulation can be integrated into existing industry QoS standards and explore its broader implications and use cases.

Intel's Resource Director Technology (RDT)~\cite{rdt}, ARM's Memory Partitioning and Monitoring (MPAM)~\cite{mpam}, and, more recently, RISC-V's Capacity/Bandwidth QoS Register Interface (CBQRI)~\cite{cbqri_latest} are industry standards for resource partitioning and bandwidth regulation. Each provide similar abstractions and enforcement interfaces for x86, ARM, and RISC-V architectures, respectively. Together, these standards illustrate the growing architectural support for hardware-assisted cache and memory QoS across modern multicore systems---a welcome development for real-time systems.

However, when it comes to memory bandwidth regulation, none of these standards provide abstractions for controlling individual DRAM banks. This is problematic for real-time systems, where worst-case performance guarantees are of paramount importance. Without bank-level control, system designers face two undesirable options: accept extremely weak worst-case guarantees or adopt heavy-handed throttling that sacrifices the high throughput potential of modern DRAM technologies, as demonstrated in this work.

Fortunately, we believe this limitation can be addressed without extensive changes to existing standards. The proposed per-bank regulation approach does not require distinct bandwidth budgets for each bank.
Thus, the existing bandwidth budget specified in these standards can simply be \textit{interpreted} as a per-bank budget rather than a system-wide budget. In other words, per-bank regulation could be realized entirely within the implementation of these standards---without modifying their architectural interfaces—provided that vendors clearly document the semantics of the bandwidth budget.

The benefits of per-bank regulation extend beyond real-time systems. For example, a recent work~\cite{Fiedler2026MemoryBandaid} proposes bandwidth regulation as a defense mechanism against RowHammer attacks~\cite{kim2014flipping}. The authors explicitly highlight the need for per-bank regulation to constrain bank-level activation rates, directly addressing the root cause of the problem while avoiding the substantial performance penalties associated with existing approaches. This example illustrates the broader utility of per-bank regulation well beyond real-time systems.
\section{Related Work} \label{sec:related}

Timing unpredictability caused by memory interference in shared-memory multicores has long been a major challenge for real-time systems. Over the past decade, the real-time systems community has extensively studied this problem. These efforts include developing memory-interference analysis models
\cite{pellizzoni2010worst,kim2014bounding,yun2015memanalysis,hassan2018bounding,hassan2020analysis},
proposing predictable memory controllers
\cite{reineke2011pret,krishnapillai2014rank,ecco2015improved,guo2017requests,guo2018comparative},
exploring DRAM bank partitioning
\cite{liu2012software,suzuki2013coordinated,yun2014palloc,pan2021numa,yang2024redb},
and bandwidth regulation mechanisms
\cite{yun2013memguard,ali2018bwlock,xu2019holistic,saeed2022utilization,zuepke2023mempol,izhbirdeev2024memcore,sun2025multi,park2025field}.

Among these efforts, memory bandwidth regulation has been one of the most actively explored approaches. Software-based mechanisms, in particular, have been popular because they are easy to deploy and widely applicable to existing COTS multicore platforms. Since MemGuard~\cite{yun2013memguard} first introduced a performance-counter-based regulation mechanism, numerous follow-up works have expanded the design space.
BWLOCK~\cite{ali2018bwlock} provided a fine-grained API for application-level bandwidth control.
Xu et al.~\cite{xu2019holistic} integrated cache partitioning and bandwidth regulation into a unified resource-allocation framework.
Saeed et al.~\cite{saeed2022utilization} proposed a dynamic regulation approach driven by DRAM-controller feedback.
MemPol~\cite{zuepke2023mempol} explored a polling-based mechanism using a dedicated real-time microcontroller.
While versatile, software-based bandwidth-regulation mechanisms suffer from relatively high overhead and limited regulation granularity.

To address these limitations, researchers have also explored hardware-based bandwidth regulators.
BRU~\cite{farshchi2020bru} introduced a drop-in regulation unit that controls traffic between CPU cores and the shared LLC.
MemCoRe~\cite{izhbirdeev2024memcore} proposed a coherence-based hardware regulator for Xilinx PS+PL systems.
AXI-REALM~\cite{benz2024axi} extended bandwidth regulation to heterogeneous SoCs by controlling accelerator traffic at the AXI interconnect.

As discussed in Section~\ref{sec:discussion}, major industry players have also begun incorporating architectural support for shared-resource management, including DRAM-bandwidth regulation. Intel's RDT~\cite{rdt} was the first to provide such mechanisms, offering memory-bandwidth monitoring (MBM) and memory-bandwidth allocation (MBA). MBA exposes a set of predefined throttling levels, but their semantics are not precisely documented and have evolved across processor generations. Unfortunately, MBA has been shown to be insufficient for achieving the isolation guarantees required by real-time workloads, even at its most restrictive configuration~\cite{sohal2022closer}.

More recently, ARM's MPAM~\cite{mpam_latest} and RISC-V's CBQRI~\cite{cbqri_latest} have introduced architectural specifications with similar goals. MPAM provides additional capabilities such as minimum/maximum bandwidth limits, and CBQRI likewise enables resource tagging and bandwidth control within the RISC-V ecosystem. However, unlike Intel RDT, both MPAM and CBQRI are still in the early stages of adoption, limiting their practical impact to date. Nevertheless, once they are broadly deployed, these hardware-based mechanisms have strong potential to become viable tools for real-time systems.

However, existing software- and hardware-based regulators are generally unaware of the banked, parallel organization of modern memory systems, which are often mapped using sophisticated XOR-based schemes to distribute load and avoid hotspots~\cite{zhang2000permutation}. Therefore, as demonstrated in this work, their worst-case effectiveness is limited. Within the security community, the importance of banked memory organization has long been recognized because of its security implications, such as Rowhammer~\cite{kim2014flipping,Fiedler2026MemoryBandaid}. Consequently, many reverse-engineering tools have been proposed and used~\cite{pessel2016drama,wi2025sudoku,heckel2025verifying}, although most of these work has focused on x86-based Intel and AMD platforms. Within the real-time systems community, reverse engineering of DRAM bank mapping on ARM-based platforms has gained prominence for analyzing and mitigating memory contention~\cite{stevanato2024learning,bechtel2021memory}. However, most prior efforts on ARM platforms have assumed simple direct mappings. In contrast, our enhancements to DRAMA~\cite{pessel2016drama} enable reverse engineering of DRAM bank mappings on both x86 and ARM platforms and have successfully uncovered sophisticated XOR-based bank-mapping schemes, thereby enabling bank-aware contention analysis and mitigation.

We believe that our proposed per-bank bandwidth-regulation approach can be implemented in a manner compatible with these existing architectural standards, and that it offers significant performance benefits. Broader adoption of architectural QoS mechanisms like the one we propose here will be crucial for enabling predictable and high-performance real-time systems on future multicore and heterogeneous platforms.
\section{Conclusion} \label{sec:conclude}

In this work, we presented a per-bank memory bandwidth regulator that enables predictable and high-performance real-time systems. Using improved DRAM bank-mapping reverse-engineering tools, we characterized the guaranteed bandwidth of modern platforms and showed that worst-case DRAM contention remains consistent across DDR generations. Building on this insight, we demonstrated a single-bank contention attack, revealing that interference peaks when attackers target the same bank and that higher aggregate bandwidth does not necessarily increase contention. Motivated by these findings, we designed and implemented a per-bank bandwidth regulator that enforces domain isolation while preserving system throughput. Our evaluation demonstrates strong temporal isolation and a 5.74$\times$ average throughput improvement over all-bank regulation. Overall, per-bank regulation is an effective and practical strategy for mitigating interference in shared DRAM systems.
\section*{Acknowledgments}\label{sec:acknowledge}
This research is supported in part by NSF grants CPS-2038923 and CCF-2403013. Connor Rudy Sullivan is supported by the Madison and Lila Self Graduate Fellowship at the University of Kansas. 

\bibliographystyle{IEEEtran}
\bibliography{references}

\end{document}